\documentstyle[12pt]{article}

\begin{document}
\centerline{{\Large{\bf The Quantum Stress-Tensor in} }}
\smallskip
\centerline{{\Large{\bf Self-Similar Spherical Dust Collapse} }}
\bigskip
\centerline{SUKRATU BARVE\footnote{E-mail address:
sukkoo@relativity.tifr.res.in }, 
T. P. SINGH\footnote{E-mail address: tpsingh@tifrc3.tifr.res.in }}
 
\centerline{{\small{\it Department of Astronomy and Astrophysics,}}}
\centerline{{\small{\it Tata Institute of Fundamental Research,}}}
\centerline{{\small{\it Homi Bhabha Road, 
Mumbai 400 005, India.}}}
\smallskip
\centerline{CENALO VAZ\footnote{E-mail address: cvaz@mozart.si.ualg.pt}}

\centerline{{\small {\it Unidade de Ci\^encias Exactas e Humanas,}}}
\centerline{{\small{\it Universidade do
Algarve,}}}
\centerline{{\small{\it Campus de Gambelas, P-8000 Faro, Portugal.}}}
\smallskip
\centerline{and}
\smallskip
\centerline{LOUIS WITTEN\footnote {E-mail address: witten@physics.uc.edu}}
\centerline{
{\small{\it Department of
Physics,}}}
\centerline{{\small{\it University of Cincinnati,}}}
\centerline{{\small{\it Cincinnati, OH 45221-0011, U. S. A.}}}

\date{}

\bigskip
\centerline{ABSTRACT}
\smallskip
{\small
\noindent We calculate the quantum stress tensor for
a massless scalar field in the 2-d self-similar spherical dust collapse
model which admits a naked singularity. We find that the outgoing radiation
flux diverges on the Cauchy horizon. This may have two consequences. 
The resultant back reaction may prevent the naked singularity from forming,
thus preserving cosmic censorship through quantum effects. The divergent 
flux may lead to an observable signature differentiating naked singularities 
from black holes in astrophysical observations.}

\newpage

\section{INTRODUCTION}

A proof or disproof of the cosmic censorship hypothesis remains an important
unsolved problem in classical general relativity. In physical terms, the
hypothesis states that the singularities arising in the gravitational
collapse of physically reasonable matter, starting from generic initial
conditions, are not visible to an outside observer. Various model examples 
studied in recent years,
especially in spherical collapse, show that either  black holes or naked
singularities may form during gravitational collapse (for recent reviews see 
\cite{ref}). While these models are not of a general enough nature to
establish that the hypothesis does not hold, it still appears reasonable to
start enquiring what the physical and possibly observational consequences of
these naked singularities could be.

Although the censorship hypothesis is framed within classical general
relativity, it is natural to expect that if the classical theory does
predict a naked singularity, quantum effects will play a fundamental role in
the extremely high curvature regions that are exposed near a naked
singularity. When the curvatures approach Planck scales, the final fate of
the collapse will be determined by quantum gravitational effects. However,
even before the Planck epoch is reached, semiclassical processes (e.g.
Hawking radiation and vacuum polarisation) will affect the formation of the
classical naked singularity. Unlike the quantum gravitational features, at
least some of the semiclassical phenomena (corresponding to quantum theory
on a classical curved background) are calculable, and these are the subject
of the present paper.

Since not much is known about cosmic censorship, quantization of matter
fields in the background of a classical dynamical spacetime containing a
naked singularity has received only sparse attention. 
Two early works which have inspired some of the recent developments in this
regard are those by Ford and Parker \cite{fp}, and by Hiscock et al. \cite
{his}. Ford and Parker considered the quantization of a massless scalar
field in a four dimensional background and gave an expression for the power
radiated to infinity in the geometric optics approximation. One of the
spacetimes they considered is the collapse of a dust cloud leading to the
formation of a shell-crossing naked singularity. They found in this
spacetime that the energy flux of the created scalar particles remains
finite up to the time of formation of the naked singularity. They also
considered the collapse of charged shells leading to naked singularities,
and for this spacetime the flux of created particles is infinite.

Hiscock et al. studied quantum effects in the classical background of a
collapsing self-similar spherical null dust cloud. This spacetime is
described by the Vaidya metric, and admits a naked singularity if the rate
of infall of the null dust is below a critical value. For the case in which
the Cauchy horizon coincides with the event horizon, they could compute the
spectrum of created particles, and the spectrum was found to be non-thermal.
For the case in which the two horizons do not coincide, they performed an
exact calculation of the vacuum stress tensor in
two dimensions and found it to diverge along the Cauchy horizon in a
positive fashion. Such a divergence is a likely indicator that the
back-reaction of the metric will prevent the formation of the naked
singularity.

Calculations such as the above can be performed only for those spacetimes
where the naked singularity formation has been demonstrated using
analytical, rather than numerical, methods. Such analytical examples are
rare. In addition to the above mentioned examples, another one which has
been studied analytically in recent years is the shell-focusing naked
singularity forming in the spherical collapse of dust \cite{dust}, described
by the Tolman-Bondi metric. Shell-focusing naked singularities are usually
regarded as more serious violations of cosmic censorship as compared to the
shell-crossing solution studied, for instance, by Ford and Parker.

In a recent work we have used the Ford and Parker geometric optics formula
to compute the radiated power in the model of self-similar dust collapse
leading to a shell-focusing naked singularity \cite{bsvw}. The assumption of
self-similarity (i.e. the existence of a homothetic Killing vector field in
the spacetime) facilitates the integration of null rays, or equivalently,
the construction of double null coordinates for the spacetime. Without such
a construction such a computation would not be possible. We find that the
radiated power diverges, in contrast to the finite flux found by Ford and
Parker in the shell-crossing case - this again appears to suggest that
shell-focusing is a more serious example of a naked singularity, as compared
to shell-crossing. We have also recently calculated the spectrum of the
radiation, and find it to differ significantly from the black body spectrum
of Hawking radiation \cite{vw}.

In the present work, we calculate the 2-d quantum
stress tensor for a minimally coupled massless scalar field in self-similar
spherical dust collapse, in a manner analogous to the calculation of Hiscock
et al. In two dimensions, the trace anomaly permits an exact calculation of
the expectation value without having to resort to geometric optics. Once
again, the assumption that the spacetime is self-similar is made. Our
results are identical to those of Hiscock et al. - we also find a divergence
of the outgoing flux on the Cauchy horizon, and the form of the divergence
is same as that for the null dust model.

The plan of the paper is as follows. In Section 2 we briefly describe the
self-similar dust collapse model. In Section 3 the standard expressions for
the quantum stress tensor in the 2-d case are recalled. In Section 4 we
construct the double null coordinates for our model and give a calculation
showing the divergence of the outgoing flux on the Cauchy horizon. In
Section 5 the detailed form of the quantum stress tensor is calculated, so
as to find exact expressions that are valid everywhere in the spacetime,
including on the Cauchy horizon.

Quantum effects have also been studied in dynamic naked singular spacetimes
in string theory inspired gravity models, and here too, features similar to
those in general relativistic collapse models (e.g. divergence of
stress-tensor on the Cauchy horizon) have been found \cite{vaz}.

\section{THE CLASSICAL SOLUTION}

The Tolman-Bondi solution represents a spherically symmetric cloud of dust
collapsing under the action of its own gravity. The energy-momentum tensor
is diagonal with the pressure terms zero and $\varepsilon $ as the energy
density. The metric is assumed to be diagonal and spherically symmetric. The
line element is given by 
\begin{equation}
\label{sm}ds^2=dt^2-e^{\lambda (r,t)}dr^2-R(r,t)^2(d\theta ^2+\sin ^2\theta
d\phi ^2) 
\end{equation}
where $t$ is coordinate time and $r$, $\theta $ and $\phi $ are the comoving
space like polar coordinates. $R(r,t)$ is called the `area radius' as the area
of a 2-d spherical surface is $4 \pi R^2$.

The field equations simplify to 
\begin{equation}
\label{f1}4\pi \varepsilon (r,t)=\frac{M^{\prime }(r)}{R^2(r,t)R^{\prime
}(r,t)} 
\end{equation}
and 
\begin{equation}
\label{f2}\dot R^2=f(r)+\frac{2M(r)}R 
\end{equation}
where `dot' indicates partial derivative with respect to time $t$, and
`prime' indicates partial derivative with respect to radial coordinate $r$. $%
f(r)$ is defined through the relation 
\begin{equation}
\label{la}e^{\lambda (r,t)}=\frac{R^{\prime 2}}{1+f(r)}. 
\end{equation}
$M(r)$ and $f(r)$ are smooth functions with the restriction that $M(r)\geq 0$
for every $r$ and $f(r)\geq -1$ for every $r$. $M(r)$ is the mass within a
sphere and $f(r)$ resembles total energy at $r$. Hence they are called mass
and energy functions respectively.

It turns out that $r=constant$ are geodesics in this spacetime. Thus we are
using a comoving frame of reference, as mentioned above.

The evolution of the cloud according to the above equations results in the
formation of a curvature singularity and the Kretschmann scalar diverges in
the approach to the singularity. The singularity at $r=0$ (usually known as
the central singularity) is known to be a naked singularity for certain
initial data, and a covered singularity for other initial data \cite{dust}.

We will be interested in those initial conditions for which the collapse of
the cloud is self-similar, i.e. it admits a homothetic Killing vector field 
\cite{self}. It can be shown that this corresponds to choosing $%
2M(r)=\lambda r$, where $\lambda $ is a constant, and $f(r)=0$. We assume
that the central singularity forms at $t=0$, and that at $t=0$ the scaling $%
R(r,0)=(3/2)^{2/3}\,\lambda ^{1/3}\,r$ holds. (Actually the scaling $R=r$ is
simpler and conventional; in our case the present choice is necessitated by
our having to extend the Tolman-Bondi coordinates in the exterior of the
cloud, as discussed in Section 4). The self-similar collapse solution is
then 
\begin{equation}
\label{sol}R=(2M(r))^{1/3}\,\left( \frac 32\left( r-t\right) \right) ^{2/3}.
\end{equation}
Collapse begins at some coordinate time $t=-t_i$, and the initial density
profile $\varepsilon (r,-t_i)\equiv \rho (r)$ can be shown to be of the
form, 
\begin{equation}
\label{ini}\rho (r)=\rho _0+\frac 1{3!}\rho _3r^3+\frac 1{6!}\rho _6r^6+...
\end{equation}
$t\frac \partial {\partial t}+r\frac \partial {\partial r}$ is the
homothetic Killing vector field. The central singularity forming in
self-similar collapse is naked for the range 
\begin{equation}
\label{ran}\frac{\lambda ^{3/2}}{12}\leq \frac{26}3-5\sqrt{3}
\end{equation}
and covered when $\lambda $ exceeds this value. The naked singularity is
locally as well as globally naked \cite{bsvw}. The derivative $R^{\prime }$
in the case of self-similar collapse will be needed subsequently, and is
given by 
\begin{equation}
\label{rp}R^{\prime }=\left( \frac{9\lambda }4\right) ^{1/3}\frac{1-z/3}{%
\left( 1-z\right) ^{1/3}}
\end{equation}
where $z=t/r$.

The geometry to the exterior of the boundary $r_b$ of the cloud is
Schwarzschild, with the Schwarzschild mass $M$ given by $M=M(r_b)=\lambda
r_b/2$. The first and the second fundamental forms for the two spacetime
metrics can be easily shown to match. The comoving time measured on the
boundary, which is the geodesic $r=r_b$, is the same independently of whether
it is thought of as comoving time of the self-similar metric or that of the
Schwarzschild exterior.

We shall work with the two-dimensional Tolman-Bondi spacetime obtained by
setting the angular part $d\Omega ^2=0$ in the spherical metric given in
Eqn. (\ref{sm}).

\section{THE 2-d QUANTUM STRESS TENSOR}

We now calculate the quantum stress tensor $<0_{out}|T_{\mu \nu }|0_{in}>$
for a minimally coupled massless scalar field $\phi $ in the background of a
2-d collapsing Tolman-Bondi dust cloud, in a manner analogous to the
calculation of Hiscock et al. \cite{his}. As is well-known, this stress
tensor exhibits a trace anomaly \cite{trace}, and the trace of $<T_{\mu \nu
}>$ is equal to ${\cal R}/24\pi ,$ where ${\cal R}$ is the Ricci scalar for
the background spacetime. Furthermore, according to the Wald axioms \cite
{wald}, the quantum stress tensor is conserved, 
\begin{equation}
\label{cons}<T^{\mu \nu }>_{;\nu }=0. 
\end{equation}

Any two-dimensional spacetime is conformally flat, and its metric can hence
be expressed using double null coordinates $\hat u$ and $\hat v$ as 
\begin{equation}
\label{glom}ds^2=C^2(\hat u,\hat v)\,d\hat u\,d\hat v. 
\end{equation}
It can be shown that the Wald axioms and the trace anomaly lead to the
following form of the quantum stress tensor in two dimensions \cite{trace} 
\begin{equation}
\label{mumu}<T_{\hat u\hat u}>=-\frac 1{12\pi }C\left( \frac 1C\right)
_{,\hat u,\hat u}+{A(\hat u),} 
\end{equation}
\begin{equation}
\label{vv}<T_{\hat v\hat v}>=-\frac 1{12\pi }C\left( \frac 1C\right) _{,\hat
v,\hat v}+{B(\hat v),} 
\end{equation}
\begin{equation}
\label{muvi}<T_{\hat u\hat v}>=\frac{{\cal R}C^2}{96\pi }. 
\end{equation}

Expressing all quantities in terms of null coordinates makes the
interpretation of $<T_{\mu \nu }>$ easier. The scalar field being massless,
its modes propagate along null rays. The functions $A(\hat v)$ and $B(\hat
v) $ are `constants' of integration which can be shown to be zero from the
following argument. We choose these null coordinates $\hat u$ and $\hat v$
according to the following prescription. The initial quantum state of the
scalar field is taken to be the standard Minkowski vacuum on ${\it I}^{-}.$
This state is also the vacuum with respect to the normal modes of the scalar
wave equation in $\hat u,\hat v$ coordinates. Since spacetime is
asymptotically flat and there is no incoming radiation from {\it I}$^{-}$ it
means that $<T_{\hat v\hat v}>$ is zero on {\it I}$^{-}$ and also that the
first term on the right hand side of (\ref{vv}) vanishes there; showing that 
$\ B(\hat v)$ is zero. Further, at the center ($r=0$) of the cloud we
require $\hat u=\hat v$ (see e.g. Section 8.1 of \cite{bd}) and $<T_{\hat
u\hat u}>=<T_{\hat v\hat v}>$, which gives $A(\hat u)$ $=0$.

\section{OUTGOING FLUX ON THE CAUCHY HORIZON}

In terms of the similarity parameter $z=t/r$, a set of double null
coordinates for the self-similar interior metric is 
\begin{equation}
\label{uv}\eta =re^{\int dz/(z-R^{\prime })},\qquad \zeta =re^{\int
dz/(z+R^{\prime })}. 
\end{equation}
In terms of these null coordinates the 2-d part of the metric (\ref{sm}) can
be written as 
\begin{equation}
\label{uv2}ds^2=\frac{e^{2\phi }}{\zeta \eta }\left( 1-\frac{R^{\prime 2}}{%
z^2}\right) \,d\zeta \,d\eta , 
\end{equation}
where $t=e^\phi $. It is understood that $z$ is an implicit function of $%
\zeta $ and $\eta $. We now want to relate these null coordinates to the
global null coordinates $\hat u$ and $\hat v$. We will assume that there
exist relations $\hat u=\hat u(\eta )$ and $\hat v=\hat v(\zeta ).$ In the
exterior of the cloud the Eddington-Finkelstein double null coordinates $u$
and $v$ for the Schwarzschild metric 
\begin{equation}
\label{sc}ds^2=\left( 1-\frac{2M}R\right) du\,dv+R^2d\Omega ^2 
\end{equation}
are given by 
\begin{equation}
\label{sc2}u=T-R^{*},\;v=T+R^{*},\quad R^{*}=R+2M\ln \left( \frac
R{2M}-1\right) . 
\end{equation}
Since the ingoing scalar field modes in Schwarzschild coordinates reduce to
the standard ingoing Minkowski modes at past null infinity, we have $\hat
v=v $. The Tolman-Bondi coordinate system can be extended to the exterior of
the cloud, and the Schwarzschild metric $(T,R)$ can be expressed in
Tolman-Bondi coordinates $(t,r)$ as follows \cite{ll}: 
\begin{equation}
\label{Tt}T=t-2\sqrt{2MR}-2M\ln \frac{\sqrt{R}-\sqrt{2M}}{\sqrt{R}+\sqrt{2M}}%
\,, 
\end{equation}
\begin{equation}
\label{Rr}R=(2M)^{1/3}\,\left( \frac 32\left( r-t\right) \right) ^{2/3}. 
\end{equation}
At the boundary $r_b$ of the cloud we get 
\begin{equation}
\label{Rrb}R=(2M)^{1/3}\,\left( \frac 32r_b\right) ^{2/3}\left( 1-z\right)
^{2/3} 
\end{equation}
and this substitution in (\ref{Tt}), along with $t=zr_b$, gives $T=T(z)$ on
the boundary. Hence $v=T(z)+R^{*}(z)=v(z)$ on the boundary. Also, 
\begin{equation}
\label{Vz}\frac{d\zeta }{dz}=\frac \zeta {z+R^{\prime }}, 
\end{equation}
giving on the boundary the relation 
\begin{equation}
\label{vV}\frac{d\hat v}{d\zeta }=\frac{dv}{d\zeta }=\frac{dv/dz}{d\zeta /dz}%
\equiv E(z,r_b). 
\end{equation}
This gives the relation between $\hat v$ and $\zeta $ on the boundary, and
the same functional relation holds inside the cloud. Hence we get that $%
d\hat v/d\zeta =B(z,r\,;r_b)$ inside the cloud. It is understood in the
above equation that $z$ is a function of $\zeta ,$ and the function $%
B(z,r\,;r_b)$ is obtained from $E(z,r_b)$ by expressing $\zeta $ in terms of 
$r$ and $z$ in the cloud. It may be noted that the relation (\ref{vV}) holds
on the boundary and is subsequently extended inside the cloud. The latter is
possible because we could write (\ref{vV}) in terms of null coordinates.

In order to relate $\hat u$ to $\eta $ we begin by noting that 
\begin{equation}
\label{UOV}\frac \zeta \eta =\frac{e^{\int dz/(z+R^{\prime })}}{e^{\int
dz/(z-R^{\prime })}}.
\end{equation}
Consider the center $r=0$. Here, for $t<0$, $z$ diverges $(z\rightarrow
-\infty )$, and the ratio on the right hand side of the above equation goes
to a constant, say $\epsilon $. ($\epsilon $ maybe equal to unity, but
that does not affect the argument). Thus, at $r=0$, $\zeta
=\epsilon \eta $. We also assume that at the center, $\hat u=\hat v$, which
gives, at $r=0$, 
\begin{equation}
\label{uU}\frac{d\hat u}{d\eta }=\epsilon \frac{d\hat v}{d\zeta }=\epsilon
B(z,r=0).
\end{equation}
Hence, at a point $(t,r)$ in the cloud the relation between $\hat u$ and $%
\eta $ is given by 
\begin{equation}
\label{uU2}\frac{d\hat u}{d\eta }=\epsilon B(z,r).
\end{equation}
Inside the cloud we can now write the metric as 
\begin{equation}
\label{uvm}ds^2=\frac{e^{2\phi }}{\zeta \eta }\left( 1-\frac{R^{\prime 2}}{%
z^2}\right) \,\frac{d\zeta }{d\hat v}\,\frac{d\eta }{d\hat u}\,d\hat ud\hat
v.
\end{equation}
We will need $\partial \phi /\partial \eta $ and $\partial z/\partial \eta $%
, and these derivatives can be had by inverting the relations (\ref{uv}). We
get 
\begin{equation}
\label{pz}\frac{\partial \phi }{\partial \eta }=\frac{z-R^{\prime }}{2z\eta }%
,\;\frac{\partial z}{\partial \eta }=\frac{z^2-R^{\prime 2}}{2\eta R^{\prime
}}.
\end{equation}
We thus have in the cloud 
\begin{equation}
\label{see}C^{2}(\hat u,\hat v)=\frac{e^{2\phi }}{\zeta \eta }\left( 1-\frac{%
R^{\prime 2}}{z^2}\right) \,\frac{d\zeta }{d\hat v}\,\frac{d\eta }{d\hat u}%
\equiv \frac{e^{2\phi }}{\zeta \eta }\psi (z,\phi )
\end{equation}
where 
\begin{equation}
\label{see2}\psi (z,\phi )=\frac{1-R^{\prime 2}/z^2}{\epsilon B^2(z,\phi )}.
\end{equation}
We can compute $\partial C/\partial \eta $ to get 
\begin{equation}
\label{cu}\frac{\partial C}{\partial \eta }=\frac {C}{2\eta} \left( \frac{%
z-R^{\prime }}z-1+\frac{z^2-R^{\prime 2}}{2R^{\prime }}\frac{\partial \psi
/\partial z}\psi +\frac{z-R^{\prime }}{2z}\frac{\partial \psi /\partial \phi 
}\psi \right) .
\end{equation}
We are interested in computing the quantum stress tensor in the approach
to the Cauchy horizon, which is given by the smaller root $z_{-}$ of $%
(z-R^{\prime })=0$ \cite{bsvw}. In this limit, $\eta $
diverges, and the function $B(z,\phi )$ and its derivatives are
well-behaved and finite. As a result, we get that as $z\rightarrow z_{-}$, 
\begin{equation}
\label{cu2}\frac{\partial C}{\partial \eta }=-\frac {C}{2\eta}
 \frac{dR^{\prime }%
}{dz},\quad \frac{\partial ^2C}{\partial \eta ^2}=\frac C{2\eta ^2}\frac{%
dR^{\prime }}{dz}\left( 1+\frac{1}{2}\frac{dR^{\prime }}{dz}\right) .
\end{equation}
The outgoing radiation flux $T_{\hat u\hat u}$, which is given by (\ref{mumu}%
), can be written as 
\begin{equation}
\label{tuu}T_{\hat u\hat u}=-\frac 1{C^2}\left( 2C_{\hat u}^2-CC_{\hat u\hat
u}\right) ,
\end{equation}
and by noting that 
\begin{equation}
\label{cuu}C_{,\hat u}=\frac{d\eta }{d\hat u}C_{,\eta }=\frac{C_{,\eta }}{%
\epsilon B},
\end{equation}
and by similarly converting $C_{,\hat u,\hat u}$ we find that in the
approach to the Cauchy horizon 
\begin{equation}
\label{tuu2}T_{\hat u\hat u}=\frac{dR^{\prime }/dz(2-dR^{\prime }/dz)}{%
4\epsilon ^2B^2\eta ^2},\quad T_{\eta \eta }=\frac{dR^{\prime
}/dz(2-dR^{\prime }/dz)}{4\eta ^2}.
\end{equation}
On the Cauchy horizon $\eta $ diverges and hence the $(\zeta ,\eta )$
coordinate system is singular, and the metric goes to zero as $(z-R^{\prime
})/\eta $. Hence we transform from $\eta $ to a new coordinate $\bar \eta $
that is well behaved on the Cauchy horizon so that in
the approach to the horizon

\begin{equation}
\label{tra}\frac{\partial \eta }{\partial \bar \eta }\equiv\frac \eta {h(\eta
)}\,\approx \,\frac \eta {z-R^{\prime }}.
\end{equation}
That such a transformation is possible is demonstrated in Section 5. In
these new coordinates $(\bar \eta ,\zeta )$ the outgoing radiation flux, in 
the approach to the Cauchy horizon, is 
\begin{equation}
\label{div}T_{\bar \eta \bar \eta }=\left( \frac{\partial \eta }{\partial
\bar \eta }\right) ^2T_{\eta \eta }\,\;\approx \,\;\frac{dR^{\prime
}/dz(2-dR^{\prime }/dz)}{4(z-R^{\prime })^2}=\frac{dR^{\prime
}/dz(2-dR^{\prime }/dz)}{4(z-z_{-})^2}.
\end{equation}
The outgoing flux $T_{\bar \eta \bar \eta }$ thus diverges on the Cauchy
horizon. Furthermore, it diverges in a positive fashion - this follows from
noting that 
\begin{equation}
\label{Rz}\frac{dR^{\prime }}{dz}=\left( \frac{4\sqrt{\lambda }}9\right)
^{2/3}\frac z{2\left( 1-z\right) ^{4/3}}>0.
\end{equation}
This quantity can be shown by a numerical check to be less than unity on the
Cauchy horizon.

We now demonstrate the divergence of the outgoing flux on the Cauchy horizon
in the Schwarzschild region outside the cloud, for which the metric is (\ref
{sc}), and $R$ is an implicit function of $u$ and $v$. We have to relate $u$
to the global coordinate $\hat u$ (we already know that $\hat v=v$). Since
we know $d\hat u/d\eta $ and $du/d\eta $, we can find $du/d\hat u$ as 
\begin{equation}
\label{lu}\frac{du}{d\hat u}=\frac{du/d\eta }{d\hat u/d\eta }.
\end{equation}
To get $du/d\eta $ explicitly we recall that $u=T-R^{*}=u(z,r_b)$ on the
boundary, from which $du/dz$ on the boundary can be calculated. Also, 
\begin{equation}
\label{uz}\frac{d\eta }{dz}=\frac \eta {z-R^{\prime }},
\end{equation}
so that 
\begin{equation}
\label{ooo}\frac{du}{d\eta }=\frac{du/dz}{d\eta /dz}=\frac{z-R^{\prime }}%
\eta \frac{du}{dz}
\end{equation}
on the boundary. Hence 
\begin{equation}
\label{oo2}\frac{du}{d\hat u}\equiv \chi (z)(z-R^{\prime })
\end{equation}
on the boundary. We assume this relation to extend in the Schwarzschild
region outside, where $z=t/r$ is related to the Schwarzschild coordinates
through the relations (\ref{Tt}) and (\ref{Rr}), and is to be thought of as
a function of $u$, by putting $r=r_b$ and imagining $t$ to be a function of $%
u$. The function $\chi (z)$ is well-behaved on the Cauchy horizon, and hence 
$du/d\hat u$ goes to zero as $(z-R^{\prime })$ on the Cauchy horizon. We may
now write the 2-d part of the Schwarzschild metric in global coordinates as 
\begin{equation}
\label{sgl}ds^2=C(\hat u,\hat v)d\hat ud\hat v=\left( 1-\frac{2M}R\right) 
\frac{du}{d\hat u}d\hat ud\hat v.
\end{equation}
It can be checked that $\partial C/\partial \hat u$ and $\partial
^2C/\partial \hat u^2$ both go to zero as $du/d\hat u$ as the Cauchy horizon
is approached. From (\ref{tuu}) it follows that $T_{\hat u\hat u}$ goes to a
finite value on the horizon. Transforming to $(u,v)$ coordinates we get 
\begin{equation}
\label{divv}T_{uu}=\left( \frac{\partial \hat u}{\partial u}\right)
^2T_{\hat u\hat u}\sim \frac 1{(z-R^{\prime })^2}\sim \frac 1{(z-z_{-})^2}
\end{equation}
showing the divergence of $T_{uu}$ on the Cauchy horizon in the
Schwarzschild region.

\section{EXACT CALCULATION OF THE QUANTUM STRESS TENSOR}

In this section we compute exact expressions for the quantum stress tensor
which are valid everywhere in the spacetime, including the approach to the
Cauchy horizon. We make the following choice of the null coordinates in the
interior, for convenience. 
\begin{equation}
\label{ano}U=\ln \eta =\ln r+\int \frac 1{z-R^{\prime }(z)}dz,\quad V=\ln
\zeta =\ln r+\int \frac 1{z+R^{\prime }(z)}dz. 
\end{equation}
We assume that the internal and external null coordinates are related by 
\begin{equation}
\label{coo}U=\alpha (u),\;v=\beta (V). 
\end{equation}
We impose the requirement that the field modes vanish at the center of the
cloud $V=U+2R_0$ where $R_0$ is a constant. Comparing the solutions to the
equations of motion at this one - dimensional locus, we obtain $\ \hat
u=\hat v$ at the center. Furthermore, on the same locus, 
\begin{equation}
\label{cen}\hat u=v=\beta (V). 
\end{equation}
As the relationship above is valid on a 1-d region, it should be valid all
over spacetime. Hence 
\begin{equation}
\label{all}\hat u=\beta (U+2R_0) 
\end{equation}
all over spacetime. Also, it is trivial to see that%
$$
\hat u=\beta (\alpha (u)+2R_0). 
$$

Having obtained all the coordinate relations one can transform the quantum
stress tensor given by Eqns. (\ref{mumu}) to (\ref{muvi}), expressing the
components in whichever coordinate system one finds convenient. We express
the components in the $(U,V)$ coordinates in the interior and $(u,v)$
coordinates in the exterior. We get 
\begin{equation}
\label{s1}<T_{UU}>=F_U(\beta ^{\prime })-F_U(A^2), 
\end{equation}
\begin{equation}
\label{s2}<T_{VV}>=F_V(\beta ^{\prime })-F_V(A^2), 
\end{equation}
and 
\begin{equation}
\label{s3}<T_{UV}>=-\frac 1{24\pi }(\ln A^2),_U,_V 
\end{equation}
where 
\begin{equation}
\label{s4}F_x(y)=\frac 1{12\pi }\sqrt{y}\left( \frac 1{\sqrt{y}}\right)
_{,x,x} 
\end{equation}
and $ds^2=A^2(U,V)\,dUdV$ is the interior Tolman-Bondi metric.

In the Schwarzschild region we get

\begin{equation}
\label{s5}<T_{uu}>=-F_u(D^2)+\alpha ^{\prime 2}F_U(\beta ^{\prime
})+F_u(\alpha ^{\prime }),
\end{equation}
\begin{equation}
\label{s6}<T_{vv}>=-F_v(D^2),
\end{equation}
\begin{equation}
\label{s7}<T_{uv}>=-\frac 1{24\pi }(\ln (D^2)),_u,_v
\end{equation}
where $ds^2=$ $D^2(u,v)\,dudv$ is the line element outside the cloud. The $%
\beta ^{\prime }$ in equations (\ref{s1}) and (\ref{s5}) is to be understood
as $\beta ^{\prime }\left( U+2R_0\right) $.

The advantage of writing the quantum stress tensor in this form is that this
allows for an easy comparison with the quantum stress tensor for black-hole
evaporation. Thus one may note that the 2-d tensor for Hawking radiation can
be written in precisely the above form (see Section 8.2 in \cite{bd}). The
various terms in the stress tensor have the following interpretation. Field
modes starting from $I^{-}$ pass through the cloud and reach $I^{+}$. While
going in, they make a contribution $F_V(\beta ^{\prime })$ to $<T_{VV}>$.
While they are outgoing, they contribute an amount $F_U(\beta ^{\prime })$
to $<T_{UU}>$. Being affected by the cloud, they contribute now even in the
Schwarzschild region as $\alpha ^{\prime }{}^2F_U(\beta ^{\prime })$. These
modes do not carry any new information about the singular boundary. Apart
from this effect, the quantum stress tensor contains vacuum polarization
terms. They are $-F_v(D^2)$, $-F_u(D^2)$, $-F_U(A^2)$ and $-F_V(A^2)$ in the
respective components of the stress-tensor. Finally, $F_u(\alpha ^{\prime })$
represents the Hawking radiation contribution.

We now obtain the explicit expressions for the quantum stress tensor in
terms of the null coordinates, all over the spacetime. Consider first the
interior metric. We define $z=1-q^3$ and calculate 
\begin{equation}
\label{a1}A^2=t^2\left[ 1-\left( \frac{9\lambda }4\right) ^{\frac 23}\left( 
\frac{2+q^3}{3q\left( 1-q^3\right) }\right) ^2\right] ,
\end{equation}
\begin{equation}
\label{a2}(A^2),_U=A^2\left[ \frac 29\left( \frac{9\lambda }4\right) ^{\frac
13}\left( \frac{q^3-1}{q^4}\right) +1\right] ,
\end{equation}
\begin{equation}
\label{a3}(A^2),_V=A^2\left[ \frac 29\left( \frac{9\lambda }4\right) ^{\frac
13}\left( \frac{1-q^3}{q^4}\right) +1\right] ,
\end{equation}
\begin{equation}
\label{a4}F_U(A^2)=\frac 1{12\pi }\left[ \frac 12\left( \frac 29\left( \frac{%
9\lambda }4\right) ^{\frac 13}\left( \frac{q^3-1}{q^4}\right) +1\right)
^2\,-\,\frac 29\left( \frac{9\lambda }4\right) ^{\frac 23}\frac{%
(q^3+2)(q^3-4)}{18q^8}\frac{A^2}{t^2-A^2}\right] ,
\end{equation}
\begin{equation}
\label{a5}F_V(A^2)=\frac 1{12\pi }\left[ \frac 12\left( \frac 29\left( \frac{%
9\lambda }4\right) ^{\frac 13}\left( \frac{1-q^3}{q^4}\right) +1\right)
^2\,-\,\frac 29\left( \frac{9\lambda }4\right) ^{\frac 23}\frac{%
(q^3+2)(q^3-4)}{18q^8}\frac{A^2}{t^2-A^2}\right] .
\end{equation}

Next, define $x$ by the relation
\begin{equation}
\label{exx}\frac{r_b-v}{2M}=\frac 2{3x^3}+\frac 2x-\frac 1{x^2}-2\ln |\frac{%
1+x}x|.
\end{equation}
$\beta ^{\prime }$ can be shown to be 
\begin{equation}
\label{bpr}
\beta ^{\prime }\left( V\right) =\frac{2M}{1+x}\left( \frac 1\lambda -\frac
2{3x^3}+\frac x\lambda +\frac 1{3x^2}\right).  
\end{equation}
Calculating $F_V(\beta')$ yields 
\begin{equation}
\label{big}F_V(\beta ^{\prime })=\frac 1{12\pi }\left[ \beta ^{\prime
2}\left( \frac{8x^7+7x^8}{16\left( 2M\right) ^2}\right) -\beta ^{\prime
}\left( \frac{4x^3}\lambda -\frac 23\right) \left( \frac{x^4\left(
1+x\right) }{8\left( 2M\right) }\right) +\frac 1{16}\left( 2+\frac{x^4}%
\lambda -\frac{2x}3\right) ^2\right] .
\end{equation}
$F_U(\beta ^{\prime })$ is the same expression but with $x$ related to the
retarded time as 
\begin{equation}
\label{x2}\frac{r_b-\beta (\alpha (u)+2R_0)}{2M}=\frac 2{3x^3}+\frac
2x-\frac 1{x^2}-2\ln |\frac{1+x}x|.
\end{equation}
It can be easily shown that $F_U(\beta ^{\prime })$ turns out to be finite
on the Cauchy horizon.

The vacuum polarization depends only on the ratio $z$, which reflects the
fact that the cloud is self-similar. 
The contribution of the modes from past infinity is finite inside the cloud.

Now, we turn to the Cauchy horizon. This is the outer homothetic Killing
horizon and turns out to be the locus $A=0$ in the interior. This presents a
technical difficulty with the coordinate system as the metric becomes
non-invertible on the Cauchy horizon. This problem is solved by transforming
to another null system ($W$ and $Z$ ) described below, and a better physical
picture is obtained.

Let $W=e^{-\tau U}$ and $Z=e^{\tau V}$ where 
\begin{equation}
\label{ta}\tau =-1+\frac{dR^{\prime }(z)}{dz}|_{z=z_{-}} 
\end{equation}
The metric component in the $W$ and $Z$ system becomes $\tau e^{\tau
(U-V)}A^2.$ From expressions for $\ U$, $V$ and $A$, this becomes

\begin{equation}
\label{t2}\tau ^2\exp \left[ \int \frac{2\tau R^{\prime }(z)}{z^2-R^{\prime
^2}(z)}dz\right] \;t^2\left( 1-\frac{R^{\prime ^2}(z)}{z^2}\right) .
\end{equation}
Expanding about the Cauchy horizon $z=z_{-}$, it becomes 
\begin{equation}
\label{t3}\tau ^2\,\left[ \frac 1{z-z_{-}}+O(z-z_{-})^0\right] \;\left[ 
\frac{2(z-z_{-})}{z_{-}}\left( 1-\frac{dR^{\prime }(z)}{dz}%
|_{z=z_{-}}\right) \,t^2+O(z-z_{-})^2\right] .
\end{equation}
This simplifies to 
\begin{equation}
\label{t5}\tau ^2\left( 1-dR^{\prime }(z)/dz|_{z=z_{-}}\right)
t^2+O(z-z_{-}).
\end{equation}
This shows that the new component is non-zero and finite at the Cauchy
horizon. Also, 
\begin{equation}
\label{t6}\frac{dU}{dW}=\frac 1r\exp \left[ -\int \frac \tau {z-R^{\prime
}(z)}dz\right] .
\end{equation}
This can be shown to be 
\begin{equation}
\label{t7}(z-z_{-})^{\tau \left( 1-dR^{\prime }(z)/dz|_{z=z_{-}}\right)
^{-1}}+O(z-z_{-}).
\end{equation}
Using the definition of $\tau $, one now obtains 
\begin{equation}
\label{t8}<T_{WW}>=\frac{-F_U(A^2)+F_U(\beta ^{\prime })}{(z-z_{-})^2}%
+O(z-z_{-}).
\end{equation}
The numerator of the first term may be shown to be finite.
The $<T_{ZZ}>$ contribution remains finite. In the $<T_{WW}>$
component, both the vacuum polarization as well as the contribution from the
modes from past infinity diverge. The divergence is inversely proportional
to the square of separation in $z$ from the horizon, a feature also observed
in collapse of a null dust shell. Thus we reproduce the results of the
previous section, while also obtaining an exact form for $<T_{\mu \nu }>$.

Consider next the exterior Schwarzschild region. We first compute the
Hawking radiation contribution $F_U(\alpha ^{\prime }{}).$ Let $w$ be
related to the advanced time coordinate as follows 
\begin{equation}
\label{e1}\frac{r_b-u}{2M}=\frac 2{3w^3}+\frac 2w+\frac 1{w^2}+2\ln |\frac{%
1-w}w|. 
\end{equation}
The retarded time relation between interior and exterior regions becomes 
\begin{equation}
\label{e2}\frac 1{\alpha ^{\prime }(u)}=\frac{2M}{1-w}\left( \frac 1\lambda
-\frac 2{3w^3}-\frac w\lambda -\frac 1{3w^2}\right) . 
\end{equation}
The expression for $F_u(\alpha ^{\prime })$ becomes

\begin{eqnarray}
\label{big2}F_u(\alpha ^{\prime })&=& \frac 1{12\pi }\left[ 
 \begin{array}{c}-\frac{\alpha
^{\prime 2}}4\left( 1-\frac{w^2}{2\lambda }+\frac w3\right) ^2+\\
\frac{\alpha
^{\prime }}{4\left( 2M\right) }\left( \frac{w^4}3+\frac{2w^8}\lambda -
\frac{2w^7}\lambda -\frac{w^5}3\right) + \\
\frac 1{16\left( 2M\right) ^2}\left(
8w^7-7w^8\right) \end{array}\right] 
\end{eqnarray}
Let us recall how the Hawking radiation result is recovered from this
expression. One can see that as $w$ tends to $1^{-}$, $u$ tends to $\infty $%
, trivially by inspecting the definition of $w$ in terms of $u$. This is the
event horizon. Naturally, one would be interested in the limit of $%
F_u(\alpha ^{\prime })$ in such a case. It is easy to see that the first and
second terms vanish in this limit. The third term becomes  $1/16\times
(2M)^{-2}$. Thus in the approach to the event horizon, 
\begin{equation}
\label{e3}F_u(\alpha ^{\prime })=\frac 1{48\pi }\times (4M)^{-2}.
\end{equation}
This is the well known inverse square dependence on Hawking temperature
which leads to the interpretation of $F_u(\alpha ^{\prime })$ as the Hawking
radiation contribution.

We show next that this expression behaves very differently when the
classical collapse ends in a naked singularity rather than a black hole. Now
there is a Cauchy horizon which lies outside the event horizon, and $u$
takes a finite value on the Cauchy horizon. As we know, the Cauchy horizon
occurs when $R^{\prime }(z)-z=0$ in the interior. This means that the time
coordinate of the intersection of the Cauchy horizon on the boundary $r=r_b$
is given by 
\begin{equation}
\label{e4}\frac t{r_b}-R^{\prime }\left( \frac t{r_b}\right) =0. 
\end{equation}
The comoving time $t$ is the same along the boundary whether taken from the
interior or Schwarzschild exterior because of matching of the first
fundamental form. Therefore, when $r=r_b$, 
\begin{equation}
\label{e5}t=\left( 1-\frac{2\lambda }{3w^3}\right) r_b. 
\end{equation}
Using this in the locus of the Cauchy horizon, one finds that 
\begin{equation}
\label{e6}\frac 1\lambda -\frac 2{3w^3}-\frac w\lambda -\frac 1{{3}w^2}=0. 
\end{equation}
This holds all over the Cauchy null ray in the exterior, as $w$ is constant
once it is known at the event of intersection with the boundary.

Going back to $F_u$, one finds that the third term is finite on the Cauchy
horizon. Apart from $\alpha ^{\prime }$, the rest of the factors in the
first and the second term are finite in the limit. $\alpha ^{\prime }$
diverges in this limit. The first term goes as $\alpha ^{\prime }{}^2$ and
hence gives the dominant divergence. To check its behavior near the Cauchy
horizon, $\alpha ^{\prime }$ is written as follows, 
\begin{equation}
\label{e7}\alpha ^{\prime }(u)=\frac{w^4-w^3}{3\left( 2m\right) }\left[ 1-%
\frac{3w^3}\lambda +\frac{3w^4}\lambda +w\right] ^{-1}.
\end{equation}
We define 
\begin{equation}
\label{e8}z_{out}(w)=1-\frac{2\lambda }{3\omega ^3}.
\end{equation}
It is physically analogous to the ratio $z=t/r$ of the interior. Then $%
\alpha ^{\prime }{}^2$ behaves like 
\begin{equation}
\label{e9}\left[ z_{out}(w)-z_{out}(w~~at~~the~~Cauchy~~horizon)\right]
^{-2}.
\end{equation}
Thus $F_u(\alpha ^{\prime })$ has an inverse square divergence in the above
sense. This occurs because of the first term in Eqn. (\ref{big2}). The
second term diverges only linearly as $\alpha ^{\prime }$

Finally, if we go back to the expression (\ref{s5}) for the outgoing flux,
we can conclude that it diverges on the Cauchy horizon. The vacuum
polarization term goes to zero at ${\it I}^{+}$. The components (\ref{s6})
and (\ref{s7}) for the stress tensor are the same as in the black hole case.

It is important to note here that the Cauchy horizon is an ordinary null ray
($u=constant$). The feature of inverse square divergence, in this case of a
function resembling the ratio $z$ of the interior (called $z_{out}(w)$), is
observed for the contribution of the modes from past infinity. The Hawking
radiation contribution goes as inverse square of the function $z_{out}$.

As far as the Cauchy horizon is concerned, if one ignores the effect of
modes from past infinity, the vacuum polarization diverges in the interior
but remains finite outside, the Hawking radiation diverging there. This
suggests that field modes originating from the singular boundary, in the
naked case, produce energy momentum on encountering a high curvature region
and transfer it to the exterior of the cloud as Hawking radiation.

Our results support our earlier 4-d flux calculation, which was carried out
in the geometric optics approximation \cite{bsvw}. The inverse square
divergence obtained is identical to what is found in the collapse of a shell
of self-similar null dust and could be generic to self-similar collapse
models. The divergence also serves to distinguish a naked singularity from a
black hole, and may be of significance if naked singularities do
occur in nature.

The divergence on the Cauchy horizon suggests that when the back-reaction of
the flux on the metric is taken into account, the formation of the naked
singularity will be avoided. This is an example of the instability of the
Cauchy horizon and a possible way of preserving cosmic censorship.

\bigskip\ 

{\noindent {\bf ACKNOWLEDGEMENTS}}

\smallskip

\noindent We acknowledge partial support of the  Junta Nacional de
Investigac\~ao Cient\'ifica e Tecnol\'ogica (JNICT) Portugal, under
contract number \break CERN/S/FAE/1172/97. C. V. and L. W. acknowledge the
partial support of NATO, under contract number CRG 920096 and L. W.
acknowledges the partial support of the U. S. Department of Energy under
contract number DOE-FG02-84ER40153. One of us (S. B.) would like to thank
Robert Wald for a useful discussion.

\end{document}